\documentclass{JHEP3}
\usepackage{amsmath,amssymb,graphics}
\usepackage{epsfig,multicol}

\title{Corrected Entropy of Friedmann-Robertson-Walker
Universe in Tunneling Method}
\author{Tao Zhu, Ji-Rong Ren, and Ming-Fan Li\\
     Institute of Theoretical Physics, Lanzhou University,
           Lanzhou 730000, P. R. China\\
    E-mail: \email{zhut05@lzu.cn},
            \email{renjr@lzu.edu.cn},
            \email{limf07@lzu.cn}}

\abstract{In this paper, we study the thermodynamic quantities of
Friedmann-Robertson-Walker (FRW) universe by using the tunneling
formalism beyond semiclassical approximation developed by
\emph{Banerjee} and \emph{Majhi}\cite{beyond0}. For this we first
calculate the corrected Hawking-like temperature on apparent horizon
by considering both scalar particle and fermion tunneling. With this
corrected Hawking-like temperature, the explicit expressions of the
corrected entropy of apparent horizon for various gravity theories
including Einstein gravity, Gauss-Bonnet gravity, Lovelock gravity,
$f(R)$ gravity and scalar-tensor gravity, are computed. Our results
show that the corrected entropy formula for different gravity
theories can be written into a general expression
(\ref{entropy-final}) of a same form. It is also shown that this
expression is also valid for black holes. This might imply that the
expression for the corrected entropy derived from tunneling method
is independent of gravity theory, spacetime and dimension of the
spacetime. Moreover, it is concluded that the basic thermodynamical
property that the corrected entropy on apparent horizon is a state
function is satisfied by the FRW universe.}


\begin{document}

\section{Introduction}
\label{secIntroduction}

Hawking radiation phenomenon of black holes shows that black holes
are not completely black, but emit thermal radiations like a black
body, with a temperature proportional to its surface gravity at the
horizon and with an entropy proportional to its horizon
area\cite{Hawking,Bekenstein}. The Hawking temperature and the
horizon entropy together with the mass of the black hole obey the
first law of thermodynamics\cite{Hawking2}. Modeling the phenomenon
of Hawking radiation one should incorporate quantum fields moving in
a background of classical gravity. Therefore quantum theory,
gravitational theory and thermodynamics meet at black holes
together. The first law of thermodynamics of black hole together
with the quantum nature of black hole lead people to consider the
connection between thermodynamics and gravity theory.

Inspired by black hole thermodynamics, \emph{Jacobson} first showed
that Einstein gravity can be derived from the fundamental
thermodynamics relation (Clausius relation) $\delta Q=TdS$ together
with the proportionality of entropy and the horizon area, presuming
that the relation holds for all Rindler causal horizons through each
spacetime point\cite{Jacobson}. With the viewpoint of
thermodynamics, Einstein equation is nothing but an equation of the
state of spacetime. Applying this idea to $f(R)$ gravity and
scalar-tensor gravity, it turns out that a non-equilibrium
thermodynamic setup has to be employed\cite{Eling,Cao-scalar}. For
another viewpoint, see\cite{Elizalde and Wu}.

\emph{Jacobson}'s derivation provides a convincing evidence for the
connection between thermodynamics and gravity theory. Recently, this
connection has been investigated extensively in literatures for
Friedmann-Robertson-Walker (FRW) universe. By assuming the apparent
horizon of FRW spacetime has an associated semiclassical
Bekenstein-Hawking entropy $S_{\texttt{BH}}$ and temperature $T_0$
\begin{eqnarray}
S_{\texttt{BH}}=\frac{A}{4\hbar},~~~~~~~~~~~T_0=\frac{\hbar}{2\pi\tilde{r}_A},
\end{eqnarray}
\emph{Cai et al}\cite{CaiJHEP2005} showed that Friedmann equations
can be derived from the first law of thermodynamics
$dE=T_0dS_{\texttt{BH}}$. Here $\hbar$ is the Planck constant, $A$
is the area of the apparent horizon, and $\tilde{r}_A$ is the radius
of the apparent horizon. Further they using the same procedure,
derived also Friedmann equations in the Gauss-Bonnet gravity and the
more general Lovelock gravity. That study has also been generalized
to $f(R)$ gravity and scalar-tensor
gravity\cite{Cao-scalar,CaiPLB2006,CaiPLB2007}. In
\cite{Cao-scalar,CaiPLB2007}, the Friedmann equations for $f(R)$
gravity and scalar-tensor gravity were derived from the first law of
thermodynamics by adding non-equilibrium corrections. In this case,
in order to construct the equilibrium thermodynamics in $f(R)$
gravity and scalar-tensor gravity, a mass-like function should be
introduced to define the energy flux crossing the apparent
horizon\cite{masslike2008}. Beside gravity theories in four
dimensions, the first law form of thermodynamics also holds on
apparent horizon in various braneworld scenarios\cite{Braneworld}.
Some other viewpoints and further developments in this direction see
\cite{FRW-thermodynamics-other} and references therein. The fact
that the first law of thermodynamics holds extensively in various
spacetime and gravity theories suggests a deep connection between
thermodynamics and gravity theory.

The thermodynamics behavior of spacetime is only one of the features
of gravity. This feature connects gravity and thermodynamics
together. Another feature is the quantum effects of spacetime, which
is related to the radiation of quantum fields from the horizon of
the spacetime. Of black holes, this radiation is usually called
Hawking radiation and was first found by Hawking\cite{Hawking}.
Hawking's original derivation of this radiation was completely based
on quantum field theory. Since then, several other derivations of
Hawking radiation were subsequently presented in literatures. Among
these derivations, a simple and physically intuitive picture is
provided by the tunneling mechanism\cite{Wilczek}. It has two
variants namely null geodesic method\cite{Wilczek} and
Hamilton-Jacobi method\cite{pada5}. The tunneling method has
attracted a lot of attention and has been applied to various black
hole spacetimes\cite{tunneling}. Among the applications of the
tunneling method, the fermion tunneling from black hole horizon has
also been investigated\cite{fermion,high black hole}. Recently, a
problem in the tunneling approach has been discussed which
corresponds to a factor two ambiguity in the original Hawking
temperature\cite{canonical}. Later, the connection between tunneling
formulism and the anomaly approach is discussed\cite{majhi}.
Recently, the derivation of Hawking black body spectrum in the
tunneling formulism is addressed\cite{majhi2} and this derivation
fills the gap in the existing tunneling formulations.

Now, inspired by the Hawking radiation of black hole spacetimes, a
question raises. That is, is there a Hawking-like radiation from the
apparent horizon of a FRW universe? In a recent
paper\cite{CaiHawk08091554}, the scalar particles' Hawking-like
radiation from the apparent horizon of a FRW universe was
investigated by using the tunneling method. Subsequently, further
investigations of the Hawking-like radiation as tunneling in a FRW
universe have been done by many
authors\cite{li,zhu2009Hawking,Chiang2008}. The calculation of the
Hawking-like radiation in tunneling method shows that a FRW universe
indeed emits particles with a physical Hawking-like temperature
$T_0=\frac{\hbar}{2\pi\tilde{r}_A}$, which is just the assumed
temperature on apparent horizon to construct the first law of
thermodynamics in FRW universe. Knowing the expression of this
Hawking-like temperature, one can apply the first law of
thermodynamics to identify explicitly the expression of the entropy
of apparent horizon in various gravity theories. However, as we have
known, when one constructs the first law of thermodynamics in FRW
universe, the expressions of the entropy for various gravity
theories are only assumptions. Thus, the tunneling method provides
an approach to directly calculate both the Hawking-like temperature
and the corresponding entropy of apparent horizon for FRW universe.

However, the tunneling method used for the Hawking-like radiation in
FRW universe is based on the semiclassical approximation. This means
that the Hawking-like temperature
$T_0=\frac{\hbar}{2\pi\tilde{r}_A}$ and the corresponding entropy
are both semiclassical results. When the completely quantum effect
is taken into account, both the Hawking-like temperature and entropy
of the apparent horizon should undergo corrections. But it is not
obvious how to go beyond this semiclassical approximation in the
tunneling method. Recently, the question that how to go beyond
semiclassical approximation in the tunneling method, have been
discussed in a series of papers by \emph{Banerjee} and \emph{Majhi}
\cite{beyond0,fermion-beyond}. And the general formalism of
tunneling beyond semiclassical approximation has been developed in
\cite{beyond0}. This formalism provides an approach to investigate
the quantum corrections to the semiclassical thermodynamic variables
of spacetime and has been studied extensively recently
\cite{beyond1,dynamics-beyond,exact-beyond1,exact-beyond2,fermion-beyond2}.
Therefore, it is of interest to investigate whether this formalism
can be generalized to the FRW universe.

In our previous work\cite{zhu2008Hawking}, by using the formalism of
tunneling beyond semiclassical approximation, we have considered the
Hawking-like radiation in a $(3+1)$-dimensional FRW spacetime. The
result yields the corrected expression of the Hawking-like
temperature and entropy of apparent horizon. Note that there is a
similar work that also treats the Hawking-like radiation to obtain
the corrected Hawking-like temperature in FRW universe via tunneling
beyond semiclassical approximation\cite{Chiang2009-beyond}. However,
the computations in \cite{zhu2008Hawking,Chiang2009-beyond} are
confined to a $(3+1)$-dimensional FRW spacetime in Einstein gravity.
The corrected expression of the entropy for generalized gravity
theories is generally not discussed. As we all have known, the first
law of thermodynamics holds not only for Einstein gravity, but also
for other gravity theories like Gauss-Bonnet gravity, Lovelock
gravity, $f(R)$ gravity and scalar-tensor gravity. Therefore, we
must ask if the tunneling formalism beyond semiclassical
approximation is still valid in investigating the quantum
corrections to the Hawking-like temperature and the corresponding
entropy of apparent horizon in generalized theories of gravity. In
the present work, we are going to investigate this problem.

In this paper, we would like to study the corrected thermodynamic
quantities of FRW universe by using the tunneling formalism beyond
semiclassical approximation. Via the tunneling calculation, we
obtain the corrected form of the Hawking-like temperature of
apparent horizon for a $(n+1)$-dimensional FRW universe. With this
corrected Hawking-like temperature, the explicit expressions of the
corrected entropy of apparent horizon for various gravity theories
including Einstein gravity, Gauss-Bonnet gravity, Lovelock gravity,
$f(R)$ gravity and scalar-tensor gravity, are computed. Our results
show that the corrected entropy formula for different gravity
theories can be written into a general expression of a same form and
this expression is also valid for black holes. This might imply that
the expression for the corrected entropy derived from tunneling
method is independent of the gravity theory, spacetime and the
dimension of the spacetime.

Therefore, the paper is organized as follows. In section
\ref{Hawking}, the tunneling of scalar particle and fermion are both
used to calculate the corrected Hawking-like temperature of apparent
horizon for a $(n+1)$-dimensional FRW universe. The derivation of
the corrected entropy for various gravity theories appear in
sections \ref{Einstein} and \ref{generalized gravity}. In section
\ref{test} we test the expression for corrected entropy in black
hole background, and section \ref{col} is left for our conclusions.

\section{Corrections to the semiclassical Hawking-like temperature}
\label{Hawking}

In this section, in order to obtain the corrected form of the
Hawking-like temperature of apparent horizon for FRW universe, we
consider both the scalar particle and fermion's Hawking-like
radiation by using the tunneling method beyond semiclassical
approximation.

For convenience of our analysis let us first begin with the standard
form of an $(n+1)$-dimensional FRW metric
\begin{eqnarray}
\label{frw}
ds^2=-dt^2+a^2(t)\left(\frac{dr^2}{1-kr^2}+r^2d\Omega_{n-1}^2
\right),
\end{eqnarray}
where $d\Omega_{n-1}^2$ denotes the line element of an
$(n-1)$-dimensional unit sphere, $a(t)$ is the scale factor of the
universe and $k$ is the spatial curvature constant which can take
values $k=+1$ (positive curvature), $k=0$ (flat), $k=-1$ (negative
curvature). Introducing $\tilde{r}=a(t)r$, the metric (\ref{frw})
can be rewritten as
\begin{eqnarray}
\label{hmetric} ds^2=h_{ab}dx^adx^b+\tilde{r}^2d\Omega_{n-1}^2,
\end{eqnarray}
where $x^a=(t,r)$ and $h_{ab}=\texttt{diag}(-1,a^2/(1-kr^2))$. In
FRW universe, there is a dynamical apparent horizon, which is the
marginally trapped surface with vanishing expansion and is defined
by the equation
\begin{eqnarray}
h^{ab}\partial_a\tilde{r}\partial_b\tilde{r}=0.
\end{eqnarray}
Using the metric (\ref{hmetric}), one can easily get the radius of
the apparent horizon for the FRW universe
\begin{eqnarray}
\tilde{r}_A=\frac{1}{\sqrt{H^2+k/a^2}},
\end{eqnarray}
where $H$ is the Hubble parameter, $H\equiv \dot{a}/a$ (the dot
represents derivative with respect to the cosmic time $t$).

In the tunneling approach of reference \cite{Wilczek} the
Painlev\'e-Gulstrand coordinates are used for the Schwarzschild
spaceime. Applying the change of radial coordinate, $\tilde{r}=ar$,
along with the above definitions of $H$ and $\tilde{r}_A$ to the
metric in \ref{frw} one obtains the Painlev\'e-Gulstrand-like metric
for FRW spacetime
\begin{eqnarray}
ds^2=-\frac{1-\tilde{r}^2/\tilde{r}_A^2}{1-k
\tilde{r}^2/a^2}dt^2-\frac{2 H \tilde{r}}{1-k \tilde{r}^2/a^2}dt
d\tilde{r}+\frac{1}{1-k\tilde{r}^2/a^2}d\tilde{r}^2+\tilde{r}^2d\Omega_{n-1}^2.\label{metric}
\end{eqnarray}
These coordinates have been used in both null geodesic method and
Hamilton-Jacobi method
\cite{CaiHawk08091554,li,zhu2008Hawking,zhu2009Hawking,Chiang2009-beyond}
to study the Hawking-like radiation from a $(3+1)$-dimensional FRW
metric.

\subsection{Scalar particle tunneling}
\label{scalar tunneling}

In this subsection we discuss scalar particle tunneling from
apparent horizon. Although there are literatures
\cite{zhu2008Hawking,Chiang2009-beyond} for the computation of the
corrections to the Hawking-like temperature via scalar particle
tunneling, they are only confined to the $(3+1)$-dimensional case.
Now, we shall do the computation for arbitrary $(n+1)$-dimensional
FRW universe.

A massless scalar field $\phi$ in FRW universe obeys the
Klein-Gordon equation
\begin{eqnarray}
\frac{-\hbar^2}{\sqrt{-g}}\partial_\mu(g^{\mu\nu}\sqrt{-g}\partial_\nu)\phi=0.\label{kg}
\end{eqnarray}
In the tunneling approach we are concerned about the radial
trajectory, so that only the $(t-\tilde{r})$ sector of the metric
(\ref{hmetric}) is relevant, thus by making the standard ansatz for
scalar wave function
\begin{eqnarray}
\phi(\tilde{r},t)=\exp\left[\frac{i}{\hbar}I(\tilde{r},t)\right],
\end{eqnarray}
the Klein-Gordon equation (\ref{kg}) can be simplified to
\begin{eqnarray}
\frac{\partial^2I}{\partial
t^2}+\left(\frac{i}{\hbar}\right)\left(\frac{\partial I}{\partial
t}\right)^2+\frac{H}{1-k\tilde{r}^2/a^2}\frac{\partial I}{\partial
t}+\frac{\tilde{r}(H^2\tilde{r}_A^2+1-k\tilde{r}^2/a^2)}{\tilde{r}_A^2(1-k\tilde{r}^2/a^2)
}\frac{\partial I}{\partial \tilde{r}}\nonumber\\-
\frac{i}{\hbar}(1-\tilde{r}^2/\tilde{r}_A^2)\left(\frac{\partial
I}{\partial
\tilde{r}}\right)^2+2\frac{i}{\hbar}H\tilde{r}\frac{\partial
I}{\partial \tilde{r}}\frac{\partial I}{\partial
t}+2H\tilde{r}\frac{\partial^2I}{\partial t\partial
\tilde{r}}-(1-\tilde{r}^2/\tilde{r}_A^2)\frac{\partial^2I}{\partial
\tilde{r}^2}=0.\label{eq1}
\end{eqnarray}
An expansion of $I(\tilde{r},t)$ in powers of $\hbar$ gives,
\begin{eqnarray}
I(\tilde{r},t)=I_0(\tilde{r},t)+\sum_i\hbar^iI_i(\tilde{r},t),\label{s}
\end{eqnarray}
where $i=1,2,3\ldots$. Substituting (\ref{s}) into (\ref{eq1}) and
equating different powers of $\hbar$ at both sides, after a
straightforward calculation we obtain the following set of
equations:
\begin{eqnarray}
\hbar^0: ~~~~~~~~~~~~~\frac{\partial I_0}{\partial
t}&=&(-H\tilde{r}\pm\sqrt{1-k\tilde{r}^2/a^2})\frac{\partial
I_0}{\partial \tilde{r}},\nonumber\\
\hbar^1: ~~~~~~~~~~~~~\frac{\partial I_1}{\partial
t}&=&(-H\tilde{r}\pm\sqrt{1-k\tilde{r}^2/a^2})\frac{\partial
I_1}{\partial \tilde{r}},\nonumber\\
\hbar^2: ~~~~~~~~~~~~~\frac{\partial I_2}{\partial
t}&=&(-H\tilde{r}\pm\sqrt{1-k\tilde{r}^2/a^2})\frac{\partial
I_2}{\partial \tilde{r}},\\
&\cdot&\nonumber\\
&\cdot&\nonumber\\
&\cdot&\nonumber\label{sets}
\end{eqnarray}
and so on. The above equations have a same functional form. So their
solutions are not independent and $I_i$ are proportional to $I_0$.
Then, we write the Eq.(\ref{s}) by
\begin{eqnarray}
I(\tilde{r},t)=(1+\sum_i\gamma_i\hbar^i)I_0(\tilde{r},t).\label{ss}
\end{eqnarray}
Here $I_0$ denotes the semiclassical contribution and the extra
value $\sum_i\gamma_i\hbar^iI_0$ can be regarded as the quantum
correction terms to the semiclassical analysis.

For the metric (\ref{metric}), since the metric coefficients are
both radius and time dependent, there is no time translation Killing
vector field as in the case of static spacetime. However, following
Kodama\cite{kodama}, for spherically symmetric dynamical spacetime
whose metric is like (\ref{metric}), there is a natural analogue,
the Kodama vector
\begin{eqnarray}
K=\sqrt{1-k\tilde{r}^2/a^2}\frac{\partial}{\partial t}.
\end{eqnarray}
Thus, using the Kodama vector, the general form of the semiclassical
action $I_0(\tilde{r},t)$ in FRW universe is given by
\begin{eqnarray}
I_0(\tilde{r},t)=-\int
\frac{\omega}{\sqrt{1-k\tilde{r}^2/a^2}}dt+\int \frac{\partial
I_0(\tilde{r},t)}{\partial \tilde{r}}d\tilde{r},\label{action0}
\end{eqnarray}
where $\omega$ is the conserved quantity with respect to the Kodama
vector $K$. The Kodama vector gives a preferred flow of time,
coinciding with the static Killing vector of standard black holes.
It should be noted that the Kodama vector is timelike, null and
spacelike as $\tilde{r}<\tilde{r}_A$, $\tilde{r}=\tilde{r}_A$ and
$\tilde{r}>\tilde{r}_A$, respectively.

Put (\ref{action0}) into the first equation of (\ref{sets}), and
combine (\ref{s}), one can obtain the solutions for
$I(\tilde{r},t)$:
\begin{eqnarray}
I(\tilde{r},t)&=&\left[-\int\frac{\omega}{\sqrt{1-k\tilde{r}^2/a^2}}dt+\omega\int
\frac{-H\tilde{r}\pm\sqrt{1-k\tilde{r}^2/a^2}}{(1-\tilde{r}^2/\tilde{r}_A^2)
\sqrt{1-k\tilde{r}^2/a^2}}d\tilde{r}\right]\nonumber\\
&&\times \left(1+\sum_i\gamma_i\hbar^i\right),\label{out}
\end{eqnarray}
where the $+(-)$ sign indicates the particle is outgoing (ingoing).

In Schwarzschild black hole, with the tunneling of a particle across
the event horizon the nature of the time coordinate $t$ changes.
This change indicates \cite{time} that $t$ coordinate has an
imaginary part for the crossing of the horizon of the black hole and
consequentially there will be a temporal contribution to the
imaginary part of the action for the ingoing and outgoing particles.
For FRW universe, the radiation is observed by a Kodama observer and
the Kodama vector is timelike, null and spacelike for the regions
outside, on and inside the apparent horizon, respectively. Because
the energy of the particle is defined by the conserved quantity with
respect to the Kodama vector, a discrepancy of Kodama vector inside
and outside the horizon will effect the temporal part of the action.
This means that the temporal part integral in (\ref{out}) should
also have an imaginary part. Therefore, outgoing and ingoing
probabilities are given by
\begin{eqnarray}
P_{\texttt{out}}&=&|\phi_{\texttt{out}}|^2=\left|\exp\left[\frac{i}{\hbar}I_{\texttt{out}}(\tilde{r},t)\right]\right|^2\nonumber\\
&=&\exp\bigg[-\frac{2}{\hbar}(1+\sum_{i}\gamma_i\hbar^i)\bigg(-\texttt{Im}\int\frac{\omega}{\sqrt{1-k\tilde{r}^2/a^2}}dt\nonumber\\
&&+\omega
\texttt{Im}\int\frac{-H\tilde{r}+\sqrt{1-k\tilde{r}^2/a^2}}{(1-\tilde{r}^2/\tilde{r}_A^2)
\sqrt{1-k\tilde{r}^2/a^2}}d\tilde{r}\bigg)\bigg],\\
P_{\texttt{in}}&=&|\phi_{\texttt{in}}|^2=\left|\exp\left[\frac{i}{\hbar}I_{\texttt{in}}(\tilde{r},t)\right]\right|^2\nonumber\\
&=&\exp\bigg[-\frac{2}{\hbar}(1+\sum_{i}\gamma_i\hbar^i)\bigg(-\texttt{Im}\int\frac{\omega}{\sqrt{1-k\tilde{r}^2/a^2}}dt\nonumber\\
&&+\omega
\texttt{Im}\int\frac{-H\tilde{r}-\sqrt{1-k\tilde{r}^2/a^2}}{(1-\tilde{r}^2/\tilde{r}_A^2)
\sqrt{1-k\tilde{r}^2/a^2}}d\tilde{r}\bigg)\bigg].
\end{eqnarray}
In \cite{zhu2009Hawking}, the temporal part contribution to the
action has been calculated in Schwarzschild-like coordinates of a
FRW spacetime. The contribution of the temporal part of the action
to the tunneling rate is canceled out when dividing the outgoing
probability by the ingoing probability because the temporal part is
completely the same for both the outgoing and ingoing solutions. It
is no need to work out the result of the temporal part of the
action.

In the WKB approximation, the tunneling probability is related to
the imaginary part of the action as
\begin{eqnarray}
\Gamma\propto\frac{P_{\texttt{in}}}{P_{\texttt{out}}}
=\exp\left[\frac{4\omega}{\hbar}\big(1+\sum_i\gamma_i\hbar^i\big)\texttt{Im}\int
\frac{1}{(1-\tilde{r}^2/\tilde{r}_A^2)}d\tilde{r}\right].
\end{eqnarray}
It is obvious that the integral function has a pole at the apparent
horizon. Through a contour integral, the tunneling probability of
ingoing particle now reads
\begin{eqnarray}
\Gamma\propto\exp\left[-\frac{2}{\hbar}\big(1+\sum_i\gamma_i\hbar^i\big)\pi
\omega\tilde{r}_A\right].
\end{eqnarray}
Now using the principle of ``detailed balance''\cite{pada5},
\begin{eqnarray}
\Gamma\propto\exp\left(-\frac{\omega}{T}\right),
\end{eqnarray}
the corrected Hawking-like temperature associated with the apparent
horizon can be determined as
\begin{eqnarray}
T=\frac{\hbar}{2\pi \tilde{r}_A}\left(1+\sum_i
\gamma_i\hbar^i\right)^{-1}=T_0\left(1+\sum_i
\gamma_i\hbar^i\right)^{-1},\label{tem}
\end{eqnarray}
where $T_0$ is the semiclassical Hawking-like temperature and other
terms are corrections coming from the higher order quantum effects.

\subsection{Fermion tunneling}
Recently, the tunneling of fermions beyond semiclassical approximation
has been also investigated for black holes\cite{fermion-beyond}. Due to the fermion tunneling
beyond semiclassical approximation, all the quantum corrrections to the
thermodynamics quantities of a black hole can be determined. In this subsection
we turn to consider the fermion tunneling beyond semiclassical approximation in FRW universe. For
the fermion tunneling, there is a paper which discusses this issue
but only with the semiclassical computation\cite{li}. Here we shall
do the analysis for the tunneling of massless fermions from a FRW
universe by considering all the quantum corrections.

Now we calculate the fermions' Hawking-like radiation from the
apparent horizon of a FRW universe via the tunneling formalism
beyond semiclassical approximation. A massless spinor field $\psi$
obeys the Dirac equation without a mass term
\begin{eqnarray}
-i\hbar\gamma^\mu D_\mu\psi=0,\label{dirac}
\end{eqnarray}
where the covariant derivative $D_\mu$ is given by
\begin{eqnarray}
D_\mu=\partial_\mu+\frac{i}{2}\Gamma_{~\mu}^{\alpha~\beta}\Sigma_{\alpha\beta},~~~~
\Sigma_{\alpha\beta}=\frac{i}{4}[\gamma^\alpha,\gamma^\beta],
\end{eqnarray}
and the gamma matrices satisfy the condition that
\begin{eqnarray}
\{\gamma^\alpha,\gamma^\beta\}=2g^{\alpha\beta}I.
\end{eqnarray}
In $(n+1)$-dimensional FRW spacetime, as in higher dimensional black
hole\cite{high black hole}, we can choose the $\gamma$ matrices for
the metric (\ref{metric}) as
\begin{eqnarray}
\gamma_{m\times m}^t&=&\left(
                       \begin{array}{cc}
                         iI_{\frac{m}{2}\times\frac{m}{2}} & 0 \\
                         0 & -iI_{\frac{m}{2}\times\frac{m}{2}} \\
                       \end{array}
                     \right),
\\
\gamma_{m\times m}^{\tilde{r}}&=&H \tilde{r}\left(
                       \begin{array}{cc}
                         iI_{\frac{m}{2}\times\frac{m}{2}} & 0 \\
                         0 & -iI_{\frac{m}{2}\times\frac{m}{2}} \\
                       \end{array}
                     \right)+\sqrt{1-k\tilde{r}^2/a^2}\left(
                               \begin{array}{cc}
                                 0 & \hat{\gamma}^3_{\frac{m}{2}\times\frac{m}{2}} \\
                                 \hat{\gamma}^3_{\frac{m}{2}\times\frac{m}{2}} & 0 \\
                               \end{array}
                             \right),
\\
\gamma_{m\times m}^\theta&=&\frac{1}{\tilde{r}}\left(
                       \begin{array}{cc}
                         0 & \hat{\gamma}^1_{\frac{m}{2}\times\frac{m}{2}} \\
                         \hat{\gamma}^1_{\frac{m}{2}\times\frac{m}{2}} & 0 \\
                       \end{array}
                     \right),
\\
\gamma_{m\times m}^\varphi&=&\frac{1}{\tilde{r}\sin\theta}\left(
                       \begin{array}{cc}
                         0 & \hat{\gamma}^2_{\frac{m}{2}\times\frac{m}{2}} \\
                         \hat{\gamma}^2_{\frac{m}{2}\times\frac{m}{2}} & 0 \\
                       \end{array}
                     \right),
\\
\dots\dots\nonumber
\\
\gamma_{m\times m}^\eta&=&\sqrt{g^{\eta\eta}}\left(
                       \begin{array}{cc}
                         0 & \hat{\gamma}^l_{\frac{m}{2}\times\frac{m}{2}} \\
                         \hat{\gamma}^l_{\frac{m}{2}\times\frac{m}{2}} & 0 \\
                       \end{array}
                     \right),~~~4\leq l\leq n
\\
\dots\dots\nonumber
\\
\gamma_{m\times m}^{x^{n+1}}&=&\sqrt{g^{x^{n+1}x^{n+1}}}\left(
                       \begin{array}{cc}
                         0 & -iI_{\frac{m}{2}\times\frac{m}{2}} \\
                         iI_{\frac{m}{2}\times\frac{m}{2}} & 0 \\
                       \end{array}
                     \right),\label{gamman}
\end{eqnarray}
where $t=x^0$ and $\tilde{r}=x^3$ are time coordinate and radial
coordinate respectively; $\theta=x^1$ and $\varphi=x^2$ are angular
coordinates, $\dots,\eta,\dots,x^{n+1}$ are extra-dimensional
coordinates; $m=2^{\frac{n+1}{2}} (m=2^{\frac{n}{2}})$ is the order
of the matrix in even-(odd-)dimensional spacetime;
$I_{\frac{m}{2}\times\frac{m}{2}}$ is a unit matrix with
$\frac{m}{2}\times\frac{m}{2}$ orders;
$\hat{\gamma}^{\mu}_{\frac{m}{2}\times\frac{m}{2}}$ is the $\mu$th
gamma matrix with $\frac{m}{2}\times\frac{m}{2}$ orders in flat
spacetime. Note that Eq.(\ref{gamman}) is only necessary in
odd-dimensional spacetime.

In the tunneling approach we are concerned about the radial
trajectory, so that only the $(t-\tilde{r})$ sector of the metric
(\ref{hmetric}) is relevant, thus the Dirac equation (\ref{dirac})
can be expressed as
\begin{eqnarray}
i\gamma_{m\times m}^t\partial_t\psi+i\gamma_{m\times
m}^{\tilde{r}}\partial_{\tilde{r}}\psi+i\chi_{m\times
m}\psi=0,\label{dirac1}
\end{eqnarray}
where $\chi_{m\times m}$ is a matrix with $m\times m$ orders and
\begin{eqnarray}
\chi_{m\times m}=\frac{i}{2}&[&\gamma^t_{m\times
m}(g^{tt}\Gamma_{tt}^{\tilde{r}}+g^{t\tilde{r}}\Gamma_{t
\tilde{r}}^{\tilde{r}}-g^{\tilde{r}t}\Gamma_{tt}^{t}-g^{\tilde{r}\tilde{r}}\Gamma_{t\tilde{r}}^{t})\nonumber\\
&+&\gamma^{\tilde{r}}_{m\times
m}(g^{tt}\Gamma_{\tilde{r}t}^{\tilde{r}}+g^{t\tilde{r}}\Gamma_{\tilde{r}
\tilde{r}}^{\tilde{r}}-g^{\tilde{r}t}\Gamma_{\tilde{r}t}^{t}-g^{\tilde{r}\tilde{r}}
\Gamma_{\tilde{r}\tilde{r}}^{t})]\Sigma_{\tilde{r}t}.
\end{eqnarray}
Without loss of generality, we employ the following ansatz for
spinor field in $(n+1)$-dimensional spacetime
\begin{eqnarray}
\psi(t,\tilde{r})=\left(
       \begin{array}{c}
         A_{\frac{m}{2}\times1}(t,\tilde{r}) \\
         B_{\frac{m}{2}\times1}(t,\tilde{r}) \\
       \end{array}
     \right)
     e^{\frac{i}{\hbar}I(t,\tilde{r})},\label{ansatz}
\end{eqnarray}
where $A_{\frac{m}{2}\times1}(t,\tilde{r})$ and
$B_{\frac{m}{2}\times1}(t,\tilde{r})$ are $\frac{m}{2}\times1$
function column matrices, $I(t,\tilde{r})$ is the one particle
action which will be expanded in powers of $\hbar$. Substituting the
ansatz (\ref{ansatz}) into (\ref{dirac1}), one obtain
\begin{eqnarray}
&(&\gamma_{m\times m}^t\partial_tI+\gamma_{m\times
m}^{\tilde{r}}\partial_{\tilde{r}}I)\left(
                                     \begin{array}{c}
                                       A_{\frac{m}{2}\times1} \\
                                       B_{\frac{m}{2}\times1} \\
                                     \end{array}
                                   \right)\nonumber\\
&-&i\hbar\left[\chi_{m\times m}\left(
                                     \begin{array}{c}
                                       A_{\frac{m}{2}\times1} \\
                                       B_{\frac{m}{2}\times1} \\
                                     \end{array}
                                   \right)
+\gamma_{m\times m}^t\left(\begin{array}{c}
                                       \partial_tA_{\frac{m}{2}\times1} \\
                                       \partial_tB_{\frac{m}{2}\times1} \\
                                     \end{array}
                                   \right)
+\gamma_{m\times m}^{\tilde{r}}\left(\begin{array}{c}
                                       \partial_{\tilde{r}}A_{\frac{m}{2}\times1} \\
                                       \partial_{\tilde{r}}B_{\frac{m}{2}\times1} \\
                                     \end{array}
                                   \right)
\right]=0.\label{dirac2}
\end{eqnarray}
Since the terms in the second line of the above equation do not
involve the single particle action, they will not contribute to the
thermodynamic entities of the black hole. Therefore, we will drop
these terms. We can expand $I$, $A_{\frac{m}{2}\times1}$, and
$B_{\frac{m}{2}\times1}$ in powers of $\hbar$ as
\begin{eqnarray}
I(t,\tilde{r})=I_0(t,\tilde{r})+\sum_i\hbar^iI_i(t,\tilde{r}),\label{action-expand}\\
A_{\frac{m}{2}\times1}(t,\tilde{r})=A_{0}(t,\tilde{r})+\sum_i\hbar^iA_i(t,\tilde{r}),\label{A-expand}\\
B_{\frac{m}{2}\times1}(t,\tilde{r})=B_{0}(t,\tilde{r})+\sum_i\hbar^iB_i(t,\tilde{r}),\label{B-expand}
\end{eqnarray}
where $i=1,2,3,\dots$. In above expansions, $I_0$, $A_0$, and $B_0$
are semiclassical values, and the other higher order terms are
treated as quantum corrections. Substituting (\ref{action-expand}),
(\ref{A-expand}), and (\ref{B-expand}) into (\ref{dirac2}), and then
equating the different powers of $\hbar$ on both sides, one obtain
the following two sets of equations:
\begin{eqnarray}
\texttt{Set
I:}&~&\nonumber
\\
\hbar^0&:&~i(\partial_tI_0+H\tilde{r}\partial_{\tilde{r}}I_0)I_{\frac{m}{2}\times\frac{m}{2}}A_0
+\sqrt{1-k\tilde{r}^2/a^2}\partial_{\tilde{r}}I_0\hat{\gamma}^3_{\frac{m}{2}\times\frac{m}{2}}B_0=0,\label{dirac-semiclassical-1}
\\
~\nonumber
\\
\hbar^1&:&~i(\partial_tI_0+H\tilde{r}\partial_{\tilde{r}}I_0)I_{\frac{m}{2}\times\frac{m}{2}}A_1
+i(\partial_tI_1+H\tilde{r}\partial_{\tilde{r}}I_1)I_{\frac{m}{2}\times\frac{m}{2}}A_0\nonumber\\
&&+\sqrt{1-k\tilde{r}^2/a^2}\partial_{\tilde{r}}I_0\hat{\gamma}^3_{\frac{m}{2}\times\frac{m}{2}}B_1
+\sqrt{1-k\tilde{r}^2/a^2}\partial_{\tilde{r}}I_1\hat{\gamma}^3_{\frac{m}{2}\times\frac{m}{2}}B_0=0,
\\
~\nonumber
\\
\hbar^2&:&~i(\partial_tI_0+H\tilde{r}\partial_{\tilde{r}}I_0)I_{\frac{m}{2}\times\frac{m}{2}}A_2
+i(\partial_tI_1+H\tilde{r}\partial_{\tilde{r}}I_1)I_{\frac{m}{2}\times\frac{m}{2}}A_1\nonumber\\
&&+i(\partial_tI_2+H\tilde{r}\partial_{\tilde{r}}I_2)I_{\frac{m}{2}\times\frac{m}{2}}A_0
+\sqrt{1-k\tilde{r}^2/a^2}\partial_{\tilde{r}}I_0\hat{\gamma}^3_{\frac{m}{2}\times\frac{m}{2}}B_2\nonumber\\
&&+\sqrt{1-k\tilde{r}^2/a^2}\partial_{\tilde{r}}I_1\hat{\gamma}^3_{\frac{m}{2}\times\frac{m}{2}}B_1
+\sqrt{1-k\tilde{r}^2/a^2}\partial_{\tilde{r}}I_2\hat{\gamma}^3_{\frac{m}{2}\times\frac{m}{2}}B_0=0,
\\
&&\dots\dots.\nonumber
\end{eqnarray}
\begin{eqnarray}
\texttt{Set II:}&~&\nonumber
\\
\hbar^0&:&~\sqrt{1-k\tilde{r}^2/a^2}\partial_{\tilde{r}}I_0\hat{\gamma}^3_{\frac{m}{2}\times\frac{m}{2}}A_0
-i(\partial_tI_0+H\tilde{r}\partial_{\tilde{r}}I_0)I_{\frac{m}{2}\times\frac{m}{2}}B_0=0,\label{dirac-semiclassical-2}
\\
~\nonumber
\\
\hbar^1&:&~\sqrt{1-k\tilde{r}^2/a^2}\partial_{\tilde{r}}I_0\hat{\gamma}^3_{\frac{m}{2}\times\frac{m}{2}}A_1
+\sqrt{1-k\tilde{r}^2/a^2}\partial_{\tilde{r}}I_1\hat{\gamma}^3_{\frac{m}{2}\times\frac{m}{2}}A_0\nonumber\\
&&-i(\partial_tI_0+H\tilde{r}\partial_{\tilde{r}}I_0)I_{\frac{m}{2}\times\frac{m}{2}}B_1
-i(\partial_tI_1+H\tilde{r}\partial_{\tilde{r}}I_1)I_{\frac{m}{2}\times\frac{m}{2}}B_0=0,
\\
~\nonumber
\\
\hbar^2&:&~\sqrt{1-k\tilde{r}^2/a^2}\partial_{\tilde{r}}I_0\hat{\gamma}^3_{\frac{m}{2}\times\frac{m}{2}}A_2
+\sqrt{1-k\tilde{r}^2/a^2}\partial_{\tilde{r}}I_1\hat{\gamma}^3_{\frac{m}{2}\times\frac{m}{2}}A_1\nonumber\\
&&+\sqrt{1-k\tilde{r}^2/a^2}\partial_{\tilde{r}}I_2\hat{\gamma}^3_{\frac{m}{2}\times\frac{m}{2}}A_0
-i(\partial_tI_0+H\tilde{r}\partial_{\tilde{r}}I_0)I_{\frac{m}{2}\times\frac{m}{2}}B_2\nonumber\\
&&-i(\partial_tI_1+H\tilde{r}\partial_{\tilde{r}}I_1)I_{\frac{m}{2}\times\frac{m}{2}}B_1
-i(\partial_tI_2+H\tilde{r}\partial_{\tilde{r}}I_2)I_{\frac{m}{2}\times\frac{m}{2}}B_0=0,
\\
&&\dots\dots.\nonumber
\end{eqnarray}

Equations (\ref{dirac-semiclassical-1}) and
(\ref{dirac-semiclassical-2}) are the semiclassical Hamilton-Jacobi
equations for a Dirac particle. Similar to the scalar particle
tunneling here one can also separate the semiclassical action as
Eq.(\ref{action0}). Substituting (\ref{action0}) into
(\ref{dirac-semiclassical-1}) and (\ref{dirac-semiclassical-2}), one
can obtain
\begin{eqnarray}
\left\{
  \begin{array}{ll}
    i\left(-\frac{\omega}{\sqrt{1-k\tilde{r}^2/a^2}}+H\tilde{r}\partial_{\tilde{r}}I_0\right)I_{\frac{m}{2}\times\frac{m}{2}}A_0
+\sqrt{1-k\tilde{r}^2/a^2}\partial_{\tilde{r}}I_0\hat{\gamma}^3_{\frac{m}{2}\times\frac{m}{2}}B_0=0,  \\
    \sqrt{1-k\tilde{r}^2/a^2}\partial_{\tilde{r}}I_0\hat{\gamma}^3_{\frac{m}{2}\times\frac{m}{2}}A_0
-i\left(-\frac{\omega}{\sqrt{1-k\tilde{r}^2/a^2}}+H\tilde{r}\partial_{\tilde{r}}I_0\right)I_{\frac{m}{2}\times\frac{m}{2}}B_0=0.
  \end{array}
\right.
\end{eqnarray}
From above equations, there will be a non-trivial solution for $A_0$
and $B_0$ if and only if the determinant of the coefficient matrix
vanishes, which results
\begin{eqnarray}
iA_0=\pm\hat{\gamma}^3_{\frac{m}{2}\times\frac{m}{2}}B_0,~~~~~~~
\partial_{\tilde{r}}I_0=\omega\frac{H\tilde{r}\pm\sqrt{1-k\tilde{r}^2/a^2}}
{(\tilde{r}^2/\tilde{r}^2_A-1)\sqrt{1-k\tilde{r}^2/a^2}},\label{result}
\end{eqnarray}
where the $+(-)$ sign indicates the particle is ingoing (outgoing).
Thus the solution for $I_0(t,\tilde{r})$ is
\begin{eqnarray}
I_0(t,\tilde{r})=-\int\frac{\omega}{\sqrt{1-k\tilde{r}^2/a^2}}dt+\omega\int
\frac{H\tilde{r}\pm\sqrt{1-k\tilde{r}^2/a^2}}{(\tilde{r}^2/\tilde{r}_A^2-1)
\sqrt{1-k\tilde{r}^2/a^2}}d\tilde{r}.\label{solution-classical}
\end{eqnarray}
Note that in the above by only solving (\ref{dirac-semiclassical-1})
and (\ref{dirac-semiclassical-2}), we obtain the solutions of
$I_0(t,\tilde{r})$. Substituting this solution
(\ref{solution-classical}) and (\ref{result}) into the equations of
Set I and Set II and then solving them we obtain relations between
different orders in the expansion of $A_{\frac{m}{2}\times1}$ and
$B_{\frac{m}{2}\times1}$:
\begin{eqnarray}
iA_j=\pm\hat{\gamma}^3_{\frac{m}{2}\times\frac{m}{2}}B_j,
\end{eqnarray}
where $j=0,1,2,3,\dots$. Using above relations about the equations
of Set I and Set II, we get the simplified form of the equations of
$I_j$:
\begin{eqnarray}
\partial_tI_j=(-H\tilde{r}\pm\sqrt{1-k\tilde{r}^2/a^2})\partial_{\tilde{r}}I_j.
\end{eqnarray}
The above set of equations have the same functional form. So their
solutions are not independent and $I_i$ are proportional to $I_0$.
Thus one can write the action $I$ as
\begin{eqnarray}
I(t,\tilde{r})=\left(1+\sum_i\gamma_i\hbar^i\right)I_0(\tilde{r},t).
\end{eqnarray}
Therefore from the above equation and (\ref{solution-classical}) one
can immediately reach the solutions of the action $I$,
\begin{eqnarray}
I(\tilde{r},t)&=&\left[-\int\frac{\omega}{\sqrt{1-k\tilde{r}^2/a^2}}dt+\omega\int
\frac{-H\tilde{r}\pm\sqrt{1-k\tilde{r}^2/a^2}}{(1-\tilde{r}^2/\tilde{r}_A^2)
\sqrt{1-k\tilde{r}^2/a^2}}d\tilde{r}\right]\nonumber\\
&&\times
\left(1+\sum_i\gamma_i\hbar^i\right),\label{fermion-solutions}
\end{eqnarray}
which is identical with the expression (\ref{out}). Following the
same step in Sec.(2.1) for scalar particle tunneling, one can obtain
the corrected Hawking-like temperature for fermion tunneling. The
result thus obtained is identical to (\ref{tem}).

\section{Corrected entropy in Einstein gravity}
\label{Einstein}

In Einstein gravity, it is known that a FRW universe can be
considered as a thermodynamical system with temperature
$T_0=\frac{\hbar}{2\pi \tilde{r}_A}$ and Bekenstein-Hawking entropy
$S_{\texttt{BH}}=\frac{A}{4\hbar}$ on the apparent horizon. Here the
temperature $T_0$ and the entropy $S_{\texttt{BH}}$ are both
semiclassical results. When the quantum effects come into play, the
temperature and the entropy should alter. In this section, with the
corrected Hawking-like temperature (\ref{tem}) on apparent horizon
of the FRW universe given in the above section, we will explicitly
calculate the corrections to the semiclassical Bekenstein-Hawking
entropy $S_{\texttt{BH}}$ with the help of the first law of
thermodynamics on the apparent horizon of the FRW universe.

In the Hawking-like temperature expression (\ref{tem}), there are
un-determined coefficients $\gamma_i$. Obviously, $\gamma_i$ should
have the dimension $\hbar^{-i}$. Now, we will perform the following
dimensional analysis to express these $\gamma_i$ in terms of
dimensionless constants by invoking some basic macroscopic
parameters of the FRW universe. In the $(n+1)$-dimensional FRW
spacetime, one sets the units as $G_{n+1}=c=k_B=1$, where $G_{n+1}$
is the $(n+1)$-dimensional gravitation constant. In this setting,
the Planck constant $\hbar$ is of the order of $m_p\cdot l_p$, where
$m_p$ is the Planck mass and $l_p$ is the Planck length. Therefore,
according to the dimensional analysis, the proportionality constants
$\gamma_i$ have the dimension of $(m_pl_p)^{-i}$. In Einstein's
gravity, there is an important quantity, i.e., the Misner-Sharp mass
\begin{eqnarray}
M=\frac{n-1}{16\pi}\Omega_{n-1}\tilde{r}^{n-2}(1-h^{ab}\partial_a\tilde{r}\partial_b\tilde{r}),
\end{eqnarray}
which is the total energy insider the sphere with radius $\tilde{r}$
and has the dimension of $m_p$. For FRW universe, the Misner-Sharp
mass on the apparent horizon is
\begin{eqnarray}
M_A=\frac{n-1}{16\pi}\Omega_{n-1}\tilde{r}_A^{n-2}.\label{MS-einstein}
\end{eqnarray}
Remember that on apparent horizon one should use
$h^{ab}\partial_a\tilde{r}\partial_b\tilde{r}\mid_{\tilde{r}_A}=0$
in the derivation of the above expression. Now we can make a
dimensional analysis to express the proportionality constants
$\gamma_i$ in terms of macroscopic parameters of the FRW universe as
\begin{eqnarray}
\gamma_i=\beta_i(M_A\tilde{r}_A)^i,\label{gama}
\end{eqnarray}
where $\beta_i$ is a dimensionless constant. This is possible since
the Misner-Sharp mass $M_A$ is independent of the radius of the
apparent horizon $\tilde{r}_A$ and can be completely determined by
$\tilde{r}_A$. Note that things will be a bit different in the next
section while in the gravity theories beyond Einstein the
Misner-Sharp mass on apparent horizon contains more than one
parameters. Using (\ref{gama}) the Hawking-like temperature
(\ref{tem}) now can be written as a new form
\begin{eqnarray}
T=T_0\left(1+\sum_i
\frac{\alpha_i\hbar^i}{(M_A\tilde{r}_A)^{i}}\right)^{-1}.\label{tem1}
\end{eqnarray}
where $\alpha_i=\frac{1}{\beta_i}$ is also a dimensionless constant.

With the new form of the Hawking-like temperature (\ref{tem1}), one
can apply the first law of thermodynamics $dE=TdS$ on apparent
horizon of the FRW universe, thus one can obtain the corrected
entropy by the formula:
\begin{eqnarray}
S=\int\frac{dE}{T}.\label{entropy}
\end{eqnarray}
substituting the temperature (\ref{tem1}) into (\ref{entropy}) we
obtain
\begin{eqnarray}
S=\int\frac{T_0}{T}dS_{\texttt{BH}}=\int \left(1+\sum_i
\frac{\alpha_i\hbar^i}{(M_A\tilde{r}_A)^{i}}\right)dS_{\texttt{BH}}.\label{entropy2}
\end{eqnarray} Note that the first law of thermodynamics with the semiclassical
temperature $T_0$ and the semiclassical Bekenstein-Hawking entropy
$S_{\texttt{BH}}$ holds on apparent horizon, i.e.,
$dE=T_0dS_{\texttt{BH}}$. In a $(n+1)$-dimensional FRW universe, the
area of the apparent horizon is $A=\Omega_{n-1}\tilde{r}_A^{n-1}$.
Thus one can obtain $M_A\tilde{r}_A=\frac{n-1}{16\pi}A$, substitute
it into (\ref{entropy2}), then we get
\begin{eqnarray}
S&=&\int \left(1+\sum_i
\alpha_i\left(\frac{4\pi}{n-1}\right)^i\frac{1}{S_{\texttt{BH}}^i}\right)dS_{\texttt{BH}}\nonumber\\
&=&S_{\texttt{BH}}+\frac{4\pi\alpha_1}{n-1}\ln
S_{\texttt{BH}}+\sum_{i=2}\frac{\alpha_i}{1-i}\left(\frac{4\pi}{n-1}\right)^i\frac{1}{S_{\texttt{BH}}^{i-1}}+\texttt{const.}\label{entropy3}
\end{eqnarray}
The first term is the semiclassical Bekenstein-Hawking entropy
$S_{\texttt{BH}}=\frac{A}{4\hbar}$. The other terms are the
correction terms due to quantum effects. Interestingly in the
correction terms, the leading order correction is logarithmic in
$S_{\texttt{BH}}$ which is very famous in black hole physics and can
be obtained by other approaches\cite{log}.

The above discussions show that in the tunneling method, the
semiclassical Bekenstein-Hawking entropy should receive corrections
due to quantum effects. In the derivation of the corrected entropy
(\ref{entropy3}), the Misner-Sharp mass and the semiclassical
Bekenstein-Hawking entropy $S_{\texttt{BH}}=\frac{A}{4\hbar}$ play
an important role. However, the Misner-Sharp mass and the entropy
for the FRW universe in Einstein gravity is very special. In
general, the Misner-Sharp mass and the entropy on the horizon in
other gravity theories are more complicated than in Einstein
gravity. Therefore, a crucial problem with the previous
investigations arises. That is, is the above procedure and the
result (\ref{entropy3}) still valid for more complicated gravity
theories? We will answer this question in the next section.

\section{Corrected entropy in generalized gravity theory}
\label{generalized gravity}

In this section, we will generalize the above discussions to the
generalized gravity theories, including the Gauss-Bonnet gravity,
Lovelock gravity, $f(R)$ gravity, and scalar-tensor gravity. We will
carry out explicitly the expression of the corrected entropy.

\subsection{Gauss-Bonnet gravity}
The  Lagrangian of the Gauss-Bonnet gravity in $(n+1)$-dimensional
spacetime is
\begin{eqnarray}
\mathcal {L}=\frac{1}{16\pi}\left(R+\alpha R_{\texttt{GB}}\right),
\end{eqnarray}
where $\alpha$ is a parameter with the dimension
$[\texttt{length}]^2$ and
$R_{\texttt{GB}}=R^2-4R_{\mu\nu}R^{\mu\nu}+R_{\mu\nu\gamma\delta}R^{\mu\nu\gamma\delta}$
is the Gauss-Bonnet term. Gauss-Bonnet gravity is the natural
generalization of Einstein gravity by including higher derivative
correction term, i.e., the Gauss-Bonnet term to the original
Einstein-Hilbert action. In this gravity theory, the semiclassical
Bekenstein-Hawking entropy-area relationship that the entropy of
horizon is proportional to its area, does not hold anymore. The
relationship is now\cite{GB-entropy}
\begin{eqnarray}
S_{\texttt{GB}}=\frac{A}{4\hbar}\left(1+\frac{n-1}{n-3}\frac{2\alpha}{r_{+}^2}\right),\label{entropy-GB-Blackhole}
\end{eqnarray}
where $A$ is the horizon area of a Gauss-Bonnet black hole and
$r_{+}$ is the radius of the horizon.

In ref.\cite{CaiJHEP2005}, \emph{Cai} et al applied the entropy
formula (\ref{entropy-GB-Blackhole}) to the apparent horizon,
assuming that the apparent horizon has an entropy with the same
expression as (\ref{entropy-GB-Blackhole}) but replacing the black
hole horizon radius $r_+$ by the apparent horizon radius
$\tilde{r}_A$. That is, the apparent horizon is supposed to have an
entropy
\begin{eqnarray}
S_{\texttt{GB}}=\frac{A}{4\hbar}\left(1+\frac{n-1}{n-3}\frac{2\alpha}{\tilde{r}_{A}^2}\right).\label{entropy-GB}
\end{eqnarray}
Then with the entropy $S_{\texttt{GB}}$ and temperature
$T_0=\frac{\hbar}{2\pi \tilde{r}_A}$ on apparent horizon, \emph{Cai}
et al have shown explicitly that the first law of thermodynamics
\begin{eqnarray}
dE=T_0dS_{\texttt{GB}}\label{first}
\end{eqnarray}
holds on apparent horizon of the FRW universe for Gauss-Bonnet
gravity, where $dE$ is the amount of energy crossing the apparent
horizon in Gauss-Bonnet gravity.

Now let us began to consider the dimensional analysis on the
Hawking-like temperature (\ref{tem1}) in Gauss-Bonnet gravity. In
Gauss-Bonnet gravity, the mass parameter is the generalized
Misner-Sharp mass, which is proposed in \cite{Misner-Sharp mass2}.
For FRW universe, the generalized Misner-Sharp mass on apparent
horizon has the following form
\begin{eqnarray}
M_A=\frac{n-1}{16\pi}\Omega_{n-1}\tilde{r}_A^{n-2}\left(1+\frac{n-1}{n-3}\frac{2\alpha}{\tilde{r}_{A}^2}\right).
\end{eqnarray}
Unlike (\ref{MS-einstein}) that has only one independent parameter
$\tilde{r}_A$, the generalized Misner-Sharp mass $M_A$ here have two
independent parameters $\tilde{r}_A$ and $\alpha$. Thus it is not
clear whether the combination $M_A\tilde{r}_A$ is still valid for
expressing the proportionality constants $\gamma_i$ in terms of
dimensionless constants. To be safe, one can express $\gamma_i$ in
terms of $\alpha$ and $\tilde{r}_A$ as
\begin{eqnarray}
\gamma_i=\beta_i(a_1
\tilde{r}_A^{n-1}+a_2\alpha\tilde{r}_A^{n-3})^i,
\end{eqnarray}
where $a_1$ and $a_2$ are dimensionless constants. Note that
$\alpha$ has the dimension $[\texttt{length}]^2$. Now the
Hawking-like temperature (\ref{tem}) has the form
\begin{eqnarray}
T=T_0\left(1+\sum_i \frac{\alpha_i\hbar^i}{(a_1
\tilde{r}_A^{n-1}+a_2\alpha\tilde{r}_A^{n-3})^i}\right)^{-1}.\label{tem-gb}
\end{eqnarray}

To fix the constants $a_1$ and $a_2$ let us first write the first
law of thermodynamics with the corrected Hawking-like temperature
(\ref{tem-gb}) in the form
\begin{eqnarray}
dS&=&\frac{dE}{T}=\frac{T_0dS_{\texttt{GB}}}{T}\nonumber\\
&=&\frac{T_0\Omega_{n-1}}{4\hbar
T}d(\tilde{r}_A^{n-1})+\frac{n-1}{n-3}\frac{T_0\Omega_{n-1}}{2\hbar
T}d(\alpha\tilde{r}_A^{n-3})\nonumber\\
&=&\frac{T_0\Omega_{n-1}}{4\hbar
T}dX+\frac{n-1}{n-3}\frac{T_0\Omega_{n-1}}{2\hbar T}dY.
\end{eqnarray}
We treat $X=\tilde{r}_A^{n-1}$, $Y=\alpha\tilde{r}_A^{n-3}$ as two
independent variables in the above equation. From the principle of
the ordinary first law of thermodynamics one interprets entropy as a
state function. In refs.\cite{exact-beyond1,exact-beyond2}, this
property of entropy has been used to investigate the first law of
thermodynamics and entropy for black holes. Also, this property must
be satisfied for FRW universe as well. Hence we can assume that the
entropy of the FRW universe is a state function and consequently
$dS$ has to be an exact differential. As a result the following
relation must hold:
\begin{eqnarray}
\frac{\partial}{\partial Y}\left(\frac{T_0\Omega_{n-1}}{4\hbar
T}\right)\bigg|_X=\frac{\partial}{\partial
X}\left(\frac{n-1}{n-3}\frac{T_0\Omega_{n-1}}{2\hbar
T}\right)\bigg|_Y.
\end{eqnarray}
This relation is just the integrability condition that ensures $dS$
is an exact differential. Using the corrected Hawking-like
temperature (\ref{tem-gb}) it follows that the above integrability
condition is satisfied only for
\begin{eqnarray}
2\frac{n-1}{n-3}a_1=a_2.
\end{eqnarray}
For convenience we choose $a_1=\frac{n-1}{16\pi}\Omega_{n-1}$, thus
the proportionality constants $\gamma_i$ is now given by
\begin{eqnarray}
\gamma_i=\beta_i\left[\frac{n-1}{16\pi}\Omega_{n-1}\tilde{r}_A^{n-2}(1+\frac{n-1}{n-3}\frac{2\alpha}{\tilde{r}_A^2})\tilde{r}_A\right]^i
=\beta_i (M_A\tilde{r}_A)^i.
\end{eqnarray}
This shows that the combination $M_A\tilde{r}_A$ still works for
expressing the proportionality constants $\gamma_i$ in terms of
dimensionless constants. Therefore the corrected form of the
Hawking-like temperature is given by
\begin{eqnarray}
T=T_0\left(1+\sum_i
\frac{\alpha_i\hbar^i}{(M_A\tilde{r}_A)^{i}}\right)^{-1}.
\end{eqnarray}

Applying the first law of thermodynamics on apparent horizon and
using (\ref{first}), one immediately obtains the corrected entropy
of apparent horizon in Gauss-Bonnet gravity
\begin{eqnarray}
S=S_{\texttt{GB}}+\frac{4\pi\alpha_1}{n-1}\ln
S_{\texttt{GB}}+\sum_{i=2}\frac{\alpha_i}{1-i}\left(\frac{4\pi}{n-1}\right)^i\frac{1}{S_{\texttt{GB}}^{i-1}}+\texttt{const},\label{entropy-GB-0}
\end{eqnarray}
which is the same in form as (\ref{entropy3}) obtained in Einstein
gravity. We see that the first term is the usual semiclassical
entropy of the horizon in Gauss-Bonnet gravity and the other terms
are the corrections from the quantum effects. Also interestingly the
leading order correction appears as the logarithmic in
$S_{\texttt{GB}}$.

Thus, starting with the Hawking-like temperature (\ref{tem1}) and
applying the first law of thermodynamics to apparent horizon, we
obtain the corrected entropy of apparent horizon in Gauss-Bonnet
gravity. The correct entropy satisfies the same formula as that in
Einstein gravity.

\subsection{Lovelock gravity}
Now we extend the above discussions to a more general case, the
Lovelock gravity, which is a generalization of the Gauss-Bonnet
gravity. The Lagrangian of the Lovelock gravity consists of the
dimensionally extended Euler densities
\begin{eqnarray}
\mathcal {L}=\sum_{i=0}^{m}c_i\mathcal {L}_i,
\end{eqnarray}
where $c_i$ are constants, $m\leq[n/2]$, and $\mathcal {L}_i$ is the
Euler density of a $(2i)$-dimensional manifold
\begin{eqnarray}
\mathcal {L}_i=2^{-i}\delta_{c_1d_1\dots c_id_i}^{a_1b_1\dots
a_ib_i}R^{c_1d_1}_{a_1b_1}\dots R_{a_ib_i}^{c^id_i}.
\end{eqnarray}
Here $\mathcal {L}_1$ is the Einstein-Hilbert term, and $\mathcal
{L}_2$ is just the Gauss-Bonnet term discussed in the previous
subsection. For the FRW universe in Lovelock gravity, the first law
of thermodynamics also holds on apparent horizon. That is
\begin{eqnarray}
dE=T_0dS_{\texttt{L}},
\end{eqnarray}
where $T_0=\frac{\hbar}{2\pi \tilde{r}_A}$ is the temperature and
$S_{\texttt{L}}$ is the entropy of the apparent horizon of the FRW
universe in Lovelock gravity, which has the following form\cite{LL-entropy}
\begin{eqnarray}
S_{\texttt{L}}=\frac{A}{4\hbar}\sum_{i=1}^{m}\frac{i(n-1)!}{(n-2i+1)!}c_i\tilde{r}_A^{2-2i}.
\end{eqnarray}

For the FRW universe in Lovelock gravity, the generalized
Misner-Sharp mass on apparent horizon is\cite{Misner-Sharp mass2}
\begin{eqnarray}
M_A=\frac{n-1}{16\pi}\Omega_{n-1}\tilde{r}_A^{n-2}\sum_{i=1}^{m}\frac{i(n-1)!}{(n-2i+1)!}c_i\tilde{r}_A^{2-2i}.
\end{eqnarray}
Now follow the same procedure in the above, we can write the
corrected Hawking-like temperature (\ref{tem1}) as the following
form
\begin{eqnarray}
T=T_0\left(1+\sum_i
\frac{\alpha_i\hbar^i}{(M_A\tilde{r}_A)^{i}}\right)^{-1}.
\end{eqnarray}
With this Hawking-like temperature we apply the first law of
thermodynamics to the apparent horizon, we can obtain the corrected
entropy
\begin{eqnarray}
S=S_{\texttt{L}}+\frac{4\pi\alpha_1}{n-1}\ln
S_{\texttt{L}}+\sum_{i=2}\frac{\alpha_i}{1-i}\left(\frac{4\pi}{n-1}\right)^i\frac{1}{S_{\texttt{L}}^{i-1}}+\texttt{const}.\label{entropy-L}
\end{eqnarray}
Also, like the corrected entropy for Gauss-Bonnet gravity, this
entropy formula follows the same form as that in Einstein gravity.

\subsection{$f(R)$ gravity}
The Lagrangian of the $f(R)$ gravity in $(n+1)$-dimensional
spacetime is
\begin{eqnarray}
\mathcal {L}=\frac{1}{16\pi}f(R),
\end{eqnarray}
where $f(R)$ is a continuous function of curvature scalar $R$. In
the $f(R)$ gravity, the entropy of a black hole has a relation to
its horizon\cite{fr-entropy}
\begin{eqnarray}
S_f=\frac{A}{4\hbar}f'(R),\label{entropy-f}
\end{eqnarray}
where $f'(R)$ denotes the derivative with respect to the curvature
scalar $R$. Also, one can assume that the apparent horizon of the
FRW universe has an entropy with the same expression as
(\ref{entropy-f}). By further assuming that the temperature
$T_0=\frac{\hbar}{2\pi\tilde{r}_A}$ still holds on the apparent
horizon, one can investigate the thermodynamics behavior of
Friedmann equations in $f(R)$ gravity. For $f(R)$ gravity, however,
things are a bit different with that in the case of Einstein
gravity, Gauss-Bonnet gravity, and Lovelock gravity. In this case,
applying the first law of thermodynamics $dE=T_0dS_f$ to the
apparent horizon, one can not obtain the correct Friedmann
equations. In order to get the correct Friedmann equations, one has
to turn from the equilibrium thermodynamics relation $dE=T_0dS_f$ to
a non-equilibrium one; an entropy production term needs to be added
to the equilibrium thermodynamics relation\cite{CaiPLB2007}. This
means that the $f(R)$ gravity corresponds to a non-equilibrium
thermodynamics of spacetime.

Recently, in \cite{masslike2008}, the authors have shown that there
is a mass-like function connecting the first law of thermodynamics
and the Friedmann equations in some gravity theories. For $f(R)$
gravity, the mass-like function can be written as
\begin{eqnarray}
\mathcal {M}=\frac{n-1}{16\pi}\Omega_{n-1}f'(R)\tilde{r}^{n-2}
(1+h^{ab}\partial_a\tilde{r}\partial_b\tilde{r}).
\end{eqnarray}
Using the mass-like function, the energy amount crossing the
apparent horizon in an infinitesimal time interval can be defined as
$dE=k^a\partial_a\mathcal {M}dt$; then the equilibrium
thermodynamics relation, i.e., the first law of thermodynamics
$dE=T_0dS_f$ holds on the apparent horizon.

The mass-like function has the dimension of $m_p$, thus we can
choose it as the mass parameter in the case of the $f(R)$ gravity.
For the FRW universe, the mass-like function on apparent horizon is
\begin{eqnarray}
\mathcal {M}_A=\frac{n-1}{16\pi}\Omega_{n-1}f'(R)\tilde{r}_A^{n-2}.
\end{eqnarray}
Thus we can write the Hawking-like temperature (\ref{tem1}) as the
form
\begin{eqnarray}
T=T_0\left(1+\sum_i \frac{\alpha_i\hbar^i}{\left(\mathcal
{M}_A\tilde{r}_A\right)^{i}}\right)^{-1}.
\end{eqnarray}
Now the procedure same as the above subsections yields the corrected
entropy formula on apparent horizon
\begin{eqnarray}
S=S_{f}+\frac{4\pi\alpha_1}{n-1}\ln
S_{f}+\sum_{i=2}\frac{\alpha_i}{1-i}\left(\frac{4\pi}{n-1}\right)^i\frac{1}{S_{f}^{i-1}}+\texttt{const}.\label{entropy-f-c}
\end{eqnarray}

\subsection{Scalar-Tensor gravity}
The general scalar-tensor theory of gravity is described by the
Lagrangian
\begin{eqnarray}
\mathcal
{L}=\frac{1}{16\pi}f(\phi)R-\frac{1}{2}g^{\mu\nu}\partial_\mu\phi\partial_\nu\phi-V(\phi),
\end{eqnarray}
where $f(\phi)$ is a continuous function of the scalar field $\phi$
and $V(\phi)$ is its potential. The black hole entropy in
scalar-tensor gravity has the following form\cite{scalar-tensor-entropy}
\begin{eqnarray}
S_{\texttt{ST}}=\frac{A}{4\hbar}f(\phi).\label{entropy-st}
\end{eqnarray}
In order to investigate the thermodynamics properties on the
apparent horizon of the FRW universe, one should assume that the
entropy of the apparent horizon has the same form as
(\ref{entropy-st}). The thermodynamics behavior in scalar-tensor
gravity is very similar with that in $f(R)$ gravity. Namely, one
usually needs to treat the thermodynamics in scalar-tensor gravity
as the non-equilibrium thermodynamics. As pointed out in
\cite{masslike2008}, after introducing the mass-like function, the
equilibrium thermodynamics $dE=T_0dS_{\texttt{ST}}$ also holds on
the apparent horizon.

The mass-like function in scalar-tensor gravity is defined as
\begin{eqnarray}
\mathcal {M}=\frac{n-1}{16\pi}\Omega_{n-1}f(\phi)\tilde{r}^{n-2}
(1+h^{ab}\partial_a\tilde{r}\partial_b\tilde{r}).
\end{eqnarray}
For the FRW universe, the mass-like function on the apparent horizon
is
\begin{eqnarray}
M_A=\frac{n-1}{16\pi}\Omega_{n-1}f(\phi)\tilde{r}_A^{n-2}.
\end{eqnarray}
Thus one can write the Hawking-like temperature as
\begin{eqnarray}
T=T_0\left(1+\sum_i \frac{\alpha_i\hbar^i}{\left(\mathcal
{M}_A\tilde{r}_A\right)^{i}}\right)^{-1}.
\end{eqnarray}
Now, with the same procedure in the above, it is easy to obtain the
corrected entropy of the apparent horizon,
\begin{eqnarray}
S=S_{\texttt{ST}}+\frac{4\pi\alpha_1}{n-1}\ln
S_{\texttt{ST}}+\sum_{i=2}\frac{\alpha_i}{1-i}\left(\frac{4\pi}{n-1}\right)^i\frac{1}{S_{\texttt{ST}}^{i-1}}+\texttt{const}.\label{entropy-st-c}
\end{eqnarray}
Obvious, this expression is consistent with the corrected entropy
formula in Einstein gravity, Gauss-Bonnet gravity, Lovelock gravity,
and $f(R)$ gravity.

It is well known that $f(R)$ gravity can be written as a special scalar-tensor theories of gravity by redefining
the field variable\cite{f-s}. To see this we writte the action of the $f(R)$ gravity as
\begin{eqnarray}\label{f-action}
I_{f(R)}=\frac{1}{16\pi}\int d^4x\sqrt{-g}f(R).
\end{eqnarray}
One can introduce a new field $\chi=R$ and write the dynamically equivalent action
\begin{eqnarray}
I_{f(R)}=\frac{1}{16\pi}\int d^4x\sqrt{-g}[f(\chi)+f'(\chi)(\chi)(\chi-R)].
\end{eqnarray}
Variation with respect to $\chi$ leads to the equation
\begin{eqnarray}
f''(R)(\chi-R)=0.
\end{eqnarray}
Therefore, $\chi=R$ if $f''(R)\neq 0$, which reproduces the action (\ref{f-action}). Then redefining the
field $\chi$ by $\phi=f'(R)$ and setting
\begin{eqnarray}
V(\phi)=\chi(\phi)\phi-f(\chi(\phi)),
\end{eqnarray}
the action takes the form
\begin{eqnarray}\label{action-BD}
I_{f(R)}=\frac{1}{16\pi}\int d^4x[\phi R-V(\phi)].
\end{eqnarray}
This is the Brans-Dicke action with a potential $V(\phi)$ and a Brans-Dicke parameter $\omega_0=0$.
Therefore there is a dynamical equivalence between $f(R)$ gravity and a special scalar-tensor gravity. This
equivalence means that the corrected entropy of FRW universe in $f(R)$ gravity can be directly obtained from
eq.(\ref{entropy-st-c}). In the gravity theory described by the action (\ref{action-BD}), the
entropy of black hole horizon takes the form $S_{\texttt{BD}}=\frac{1}{4\hbar}\phi A$. Thus from (\ref{entropy-st-c})
the corrected entropy of FRW universe can be written as
\begin{eqnarray}
S=S_{\texttt{BD}}+\frac{4\pi\alpha_1}{n-1}\ln
S_{\texttt{BD}}+\sum_{i=2}\frac{\alpha_i}{1-i}\left(\frac{4\pi}{n-1}\right)^i\frac{1}{S_{\texttt{BD}}^{i-1}}+\texttt{const},
\end{eqnarray}
noticing that $\phi=f'(R)$, which shows above expression is just the corrected entropy formula (\ref{entropy-f-c})
of the $f(R)$ gravity.

Thus, summarizing Eqs. (\ref{entropy3}), (\ref{entropy-GB-0}),
(\ref{entropy-L}), (\ref{entropy-f-c}) and (\ref{entropy-st-c}), one
can conclude that all these corrected entropy formulaes in different
gravity theories can be written into a general expression
\begin{eqnarray}
S=S_0+\tilde{\alpha}_1\ln
S_0+\sum_{i=2}\tilde{\alpha}_i\frac{1}{S_0^{i-1}}+\texttt{const},\label{entropy-final}
\end{eqnarray}
where $S_0$ is the entropy on apparent horizon without quantum
correction. This might imply that this general expression is
independent of the concrete gravity theory. Also, one can see that
the leading order correction in (\ref{entropy-final}) appears as the
logarithmic in $S_0$ and the sub-leading term is the standard
inverse power of $S_0$. This character holds for arbitrary
$(n+1)$-dimensional FRW spacetime.

\section{Test the expression for corrected entropy with black holes}
\label{test}

In above sections, it is shown that there is a general expression
for the corrected entropy on apparent horizon by the tunneling
method. However, the derivation of this general expression is
confined to the FRW universe. As we all know the tunneling method
has been used extensively to obtain the corrected black hole
entropy. Thus, a question arises that is the general expression
(\ref{entropy-final}) still valid for black holes? Now, in order to
answer this question we are going to check the corrected entropy
from the tunneling method for a $(2+1)$-dimensional BTZ black hole
and a $(3+1)$-dimensional Kerr-Newman black hole.

In ref.\cite{exact-beyond1}, \emph{Modak} has considered the
tunneling method beyond semiclassical approximation for BTZ black
hole and obtained the corresponding corrected entropy for the BTZ
black hole, which is
\begin{eqnarray}
S=S_{\texttt{BH}}+4\pi\beta_1\ln
S_{\texttt{BH}}-\frac{16\pi^2\beta_2}{l}\left(\frac{1}{S_{\texttt{BH}}}\right)+\dots,\label{btz}
\end{eqnarray}
where $S_{\texttt{BH}}=\frac{A}{4\hbar}$ is the semiclassical
Bekenstein-Hawking entropy of the BTZ black hole and $l$ is related
to a negative cosmological constant $\Lambda=-\frac{1}{l^2}$. It is
obvious that this entropy formula (\ref{btz}) fit into the general
expression (\ref{entropy-final}).

For the Kerr-Newman black hole, its corrected entropy
is\cite{exact-beyond2}
\begin{eqnarray}
S=S_{\texttt{BH}}+2\pi\beta_1\ln
S_{\texttt{BH}}-\frac{4\pi^2\beta_2\hbar}{S_{\texttt{BH}}}+\dots,\label{kn}
\end{eqnarray}
which also fits into the general expression (\ref{entropy-final}).
Now it is trivial, as one can check, that all other black holes, for
example Schwarzschild, Kerr or Reissner-Nordstrom black hole, also
fit into the general expression (\ref{entropy-final}). Thus the
universality of the expression (\ref{entropy-final}) for black holes
is justified.

Now, one can say that the general expression (\ref{entropy-final})
is also valid for black holes. Here are some comments. First, the
BTZ black hole is a black hole solution for $(2+1)$-dimensional
gravity with a negative cosmological constant, this implies that
(\ref{entropy-final}) is robust for black hole even in low
dimensional gravity theories. Second, we have noticed that
prefactors of both the logarithmic term and the third term in
(\ref{entropy-final}) are dependent on the black holes. Also, from
above section, it is easy to know that these prefactors are
dependent on the dimension of the spacetime. Although the general
expression is independent of gravity theory, spacetime and the
dimension of the spacetime in form, this means that the prefactors
may contain more detailed information of the spacetime. Third, in
the derivation of the corrected entropy (\ref{btz}) in
ref.\cite{exact-beyond1} and (\ref{kn}) in ref.\cite{exact-beyond2},
the condition that the entropy of a black hole must be a state
function is enforced. The property that entropy is a state function
is a basic character of the ordinary first law of thermodynamics.

\section{Conclusions}
\label{col}

In this paper, we have investigated the thermodynamic quantities of
FRW universe by using the tunneling formalism beyond semiclassical
approximation developed by \emph{Banerjee} and
\emph{Majhi}\cite{beyond0}. Both the scalar particle and fermion
tunneling from apparent horizon are considered to obtain the
corrected Hawking-like temperature in FRW universe. With this
corrected Hawking-like temperature, the corresponding corrected
entropy on apparent horizon for Einstein gravity, Gauss-Bonnet
gravity, Lovelock gravity, $f(R)$ gravity and scalar-tensor gravity
are given. We found that the corrected entropy formula for different
gravity theories can be written into a general expression
(\ref{entropy-final}) in form. We also show that this general
expression is valid for black holes. These characteristics may imply
that this general expression for the corrected entropy derived from
tunneling method is independent of gravity theory, spacetime and
dimension of the spacetime.

An important part in the derivation of the corrected entropy for
various gravity theories is that we have use the combination
$M_A\tilde{r}_A$ to express the proportionality constants $\gamma_i$
in terms of dimensionless constants by dimensional analysis. This
combination is always valid for Einstein gravity, but in generalized
gravity theories its validity is not clear. We have shown in
Gauss-Bonnet gravity that this combination is an essential condition
to ensure that the corrected entropy $S$ is a state function, which
is a basic property of ordinary first law of thermodynamics. This
means that the basic thermodynamical property that corrected entropy
on apparent horizon is a state function is satisfied by the FRW
universe.

There is another significant point in the general expression
(\ref{entropy-final}) for corrected entropy is that it involves
logarithmic in $S_0$ term as the leading correction together with
the standard inverse power of $S_0$ as sub-leading correction. The
prefactors of the correction terms are dependent on the dimension of
the FRW spacetime. In black holes, it has been proved that the
prefactor of logarithmic correction term is related with the trace
anomaly of the stress tensor near the horizon. In this paper, we
have not discussed this issue for the FRW universe, and thus it
remains an open issue to consider the connection between the
prefactor of logarithmic term and the trace anomaly in FRW universe
for various gravity theories.

\section*{Acknowledgements}
This work was supported by the National Natural Science Foundation
of China (No. 10275030) and Cuiying Project of Lanzhou University
(225000-582404). \emph{Tao Zhu} thanks Dr. \emph{Zhen-Bin Cao} for
helpful discussions on the dimensional analysis.

\end{document}